\documentclass[iop]{emulateapj}

\slugcomment{Accepted for publication in the Astrophysical Journal}

\shorttitle{Spitzer and near-infrared observations of a new bi-polar outflow in the RMC}
\shortauthors{Ybarra et al.}

\begin{document}


\title{Spitzer and near-infrared observations of a new bi-polar protostellar outflow in the Rosette Molecular Cloud}


\author{Jason E. Ybarra\altaffilmark{1} and Elizabeth A. Lada}
\affil{Department of Astronomy, University of Florida, Gainesville, FL 32605}
\email{jybarra@astro.ufl.edu}

\author{Zoltan Balog}
\affil{Max-Planck-Institut f\"ur Astronomie, Heidelberg, Germany}

\author{Scott W. Fleming\altaffilmark{1}}
\affil{Department of Astronomy, University of Florida, Gainesville, FL 32605}

\author{Randy L. Phelps\altaffilmark{2}}
\affil{National Science Foundation, Office of Integrative Activities, Arlington, VA 22230}
\altaffiltext{1}{Visiting astronomer, Cerro Tololo Inter-American Observatory, National Optical Astronomy Observatory, which are operated by the Association of Universities for Research in Astronomy, under contract with the National Science Foundation.}
\altaffiltext{2}{This material is based on work supported by the National Science Foundation.
Any opinion, findings, and conclusion or recommendations
expressed in this material are those of the author and do not necessarily
reflect the views of the National Science Foundation}



\begin{abstract}
We present and discuss \emph{Spitzer} and near-infrared H$_{2}$ observations of
a new bi-polar protostellar outflow in the Rosette Molecular Cloud. The outflow
is seen in all four IRAC bands and partially as diffuse emission in the MIPS 24 $\mu$m band. 
An embedded MIPS 24 $\mu$m source bisects the outflow and appears to 
be the driving source.  This source is coincident with a
dark patch seen in absorption in the 8 $\mu$m IRAC image.
\emph{Spitzer} IRAC color analysis of the shocked emission was performed from
which thermal and column density maps of the outflow were constructed. 
Narrow-band near-infrared (NIR) images of the 
flow reveal H$_2$ emission features coincident with the high temperature
regions of the outflow. This outflow has now
been given the designation MHO 1321 due to the detection of NIR H$_2$ features.
We use these data and maps to probe the physical conditions and structure of the flow.

\end{abstract}





\section{Introduction}

Outflows and jets from young stellar objects (YSOs) accompany
the early stages of star formation. Outflows can manifest themselves
as jets and knots of shocked material visible at optical and
near-infrared wavelengths and also molecular emission observable
at longer wavelengths. The outflowing material plays
a role in removing the excess angular momentum from the 
YSOs allowing them to evolve into stars. 
Outflows are able to trace the history of mass loss and accretion
of their driving sources. 
%
Studying the structure and properties of these flows may
provide clues to understanding the connection between jets and
the associated wide angle molecular flows \citep{rei2001}.
Additionally, this outflowing material
interacts with its surroundings and may affect its environment, 
possibly regulating further star formation and cluster evolution.
The energy and momentum inputted by outflows may disrupt
the surrounding ambient gas, contribute to the turbulence in the cloud,
and affect chemical processes \citep{bal2007}.
\citet{yba2009} developed a technique to study 
the thermal structure of shocked H$_2$ gas
using color analysis of observations from the \emph{Spitzer} InfraRed 
Array Camera (IRAC). 
Given the vast amount of \emph{Spitzer} data  available, this technique
can be used to survey large regions and simultaneously find and analyze
shocked emission.
The IRAC color analysis enables the construction of temperature maps of the shocked gas which may in turn be used to probe the interaction of outflow with its surroundings. 
These maps may also be used to compare 
the properties of outflow with those of simulations allowing a better 
understanding of the physics involved and estimating the
energy and momentum inputted by outflows into their environment.

The Rosette Molecular Cloud (RMC) is a star forming region located at a distance of 
1.6 kpc. Near-infrared imaging studies have revealed nine embedded clusters across
the cloud \citep{phe1997,rom2008}.  Outflow activity in the cloud has been revealed
through the [\ion{S}{2}] narrowband imaging survey of \citet{yba2004} and the 
$^{12}$CO survey of \citet{den2009}.
In an analysis of the \emph{Spitzer} IRAC images of the Rosette Molecular Cloud, 
we have discovered a
structure with the morphology of a bipolar outflow that is visible in
the images from all four IRAC bands. 
This structure can be seen in the images published by \citet{pou2008} although it 
is not discussed in their paper.

In this study we analyze the outflow using near infrared (NIR) narrowband 
imaging of the flow to confirm the presence of shocked gas 
inferred from analysis of the the IRAC images. 
We improve the IRAC color analysis of \citet{yba2009} and use it to create
temperature and column density maps of the outflow. Using both
the NIR and IRAC data, we probe the physical conditions and structure of the
outflow.

\begin{figure*}
\plotone{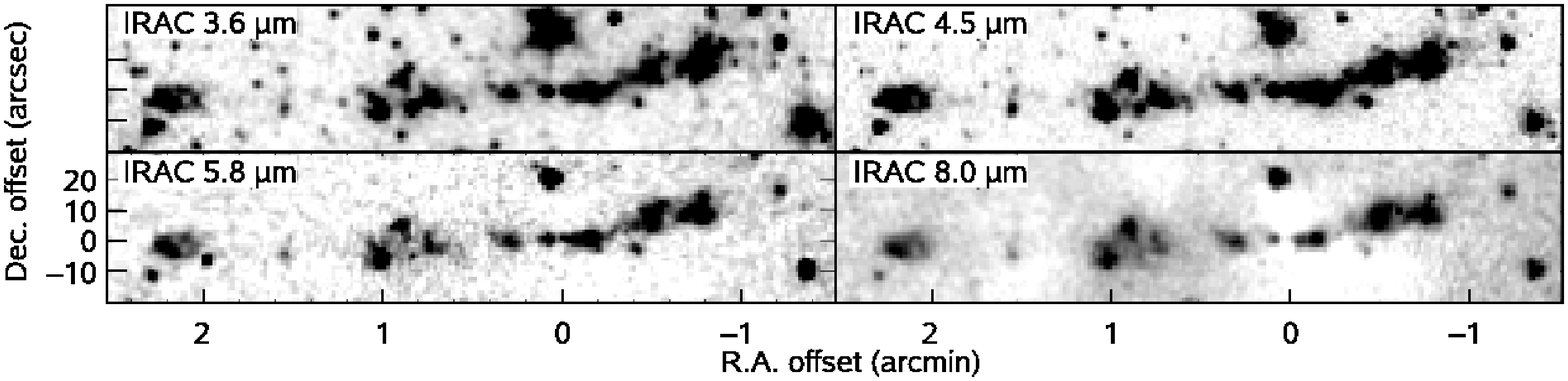}
\caption{ Spitzer IRAC images of the outflow. 
The origin is set at 
($\alpha$,$\delta$)(J2000) =
($06^{\rm{h}}35^{\rm{m}}25\fs 0$, $+03\arcdeg56\arcmin21\arcsec$)
}
\end{figure*}

\section{Observations and Data Reduction}

\subsection{Spitzer IRAC and MIPS data reduction}

We used MIPS 24$\mu$m and IRAC 3.6-8.0 $\mu$m data from program 3391
(PI: Bonnel) available in the {\it Spitzer } archive. The IRAC frames
were processed using the Spitzer Science Center (SSC) IRAC Pipeline
v14.0, and mosaics were created from the basic calibrated data (BCD)
frames using a custom IDL program (see \citet{Guter08} for details). The
MIPS frames were processed using the MIPS Data Analysis Tool
\citep{Gordon05}.

\subsection{Near-Infrared H$_2$ observations and data reduction}

Near-Infrared, narrow-band observations of the outflow were obtained
with the Infrared Side Port-Imager (ISPI) on the Blanco 4 meter
telescope at the Cerro Tololo Inter-American Observatory (CTIO).  ISPI
employs a 2048 $\times$ 2048 HgCdTe Hawaii-2 array with a 10.25
$\times$ 10.25 arcmin field of view and a plate scale of $\sim$
0.305$''$ pixel$^{-1}$. The outflow was imaged using the H$_2$ 1$-$0
S(2) 2.034 $\mu$m filter ($\lambda_{c}$ = 2.0336 $\mu$m, $\Delta \lambda/\lambda = .007$), 
H$_2$ 1$-$0 S(1) 2.122 $\mu$m filter
($\lambda_{c}$ = 2.1262 $\mu$m, $\Delta \lambda/\lambda = .01$), 
and K$_{\rm{cont}}$ filter centered at
2.1462 $\mu$m. The telescope was dithered with a 20 point dither
pattern with a integration time of 60s at each dither position. The
images were taken on the nights of 2008 December 17 and 2008 December
19 with total integration time of 40 minutes in each filter.

The raw images were flat fielded, corrected for bad pixels, and
linearized using the task {\tt{osiris}} from the CTIO Infrared
Reduction package. For each image, a sky frame was created by median
combining the dithered images closest in time to the image. 
The IRAF tasks {\tt{msctpeak}}
and {\tt{mscimage}}, which are part of the IRAF Mosaic Data Reduction
Package, {\tt{mscred}}, were used to correct the images for geometric
distortions. The high order polynomial distortion terms were
calculated with {\tt{msctpeak}} using the 2MASS Point Source Catalog
as the reference catalog and the distortion correction was applied
with {\tt{mscimage}}. The corrected images were aligned and then
combined to form the final science images.

\section{Results and Analysis}

Figure 1 shows the outflow in all four IRAC bands.  
The outflow appears as patchy regions of diffuse emission with an overall structure
that is elongated and collimated in the E-W direction.
Figure 2 shows the narrowband near-infrared emission images of the outflow. 
The NIR H$_2$ knots coincide with the diffuse emission seen in the
IRAC images.  The NIR H$_2$ images confirm the presence of shocked gas
and the interpretation that this structure is an outflow.
The eastern end appears to truncate at a bow shock. Slightly west
of the  bow shock, the H$_2$ images reveal a small scale chaotic structure followed
by a more linear chain of knots.The western end of the outflow appears slightly deflected 
northward followed by a bright knot (g) and then a complex structure of smaller knots.

The NIR images were flux calibrated to the 2MASS Ks band by
determining the magnitude difference between the 2MASS Ks band
catalog values and the ISPI image magnitudes for stars in common.
The zero point flux in  
each NIR image after calibration is then the filter bandwidth multiplied by
the 2MASS Ks-band zero point flux density. This results in a relation between
the counts sec$^{-1}$ in each filter image and the flux in W cm$^{-2}$.
The narrow band continuum K$_{\rm{c}}$ images were scaled and subsequently
subtracted from
the H$_2$ line images. Figures 2c and 2d show the continuum subtracted
H$_2$ 1--0 S(1) 2.122 $\mu$m and H$_2$ 1--0 S(2) 2.034 $\mu$m line
images.  The quality of the subtraction is good although there are some
subtraction residuals from the brightest stars present in the subtracted images due
to differences in wavelength and PSF combined with changing
atmospheric conditions.

Based on the observation, this outflow has been given the designation
MHO 1321 in the Catalogue of Molecular Hydrogen Emission-Line Objects
(MHOs) in Outflows from Young Stars\footnote{MHO catalogue is hosted by the Joint Astronomy Centre, Hawaii. \url{http://www.jach.hawaii.edu/UKIRT/MHCat/}} \citep{dav2009}. 
The fluxes of the individual
H$_2$ knots comprising this outflow were determined using a circular
aperture on the continuum subtracted images. Table 1 lists the NIR fluxes of
the H$_2$ emission knots. The flux
uncertainty is composed of the rms background, poisson noise, and
uncertainty from the flux calibration.

\notetoeditor{Please print this figure across both columns}
\begin{figure*}
\plotone{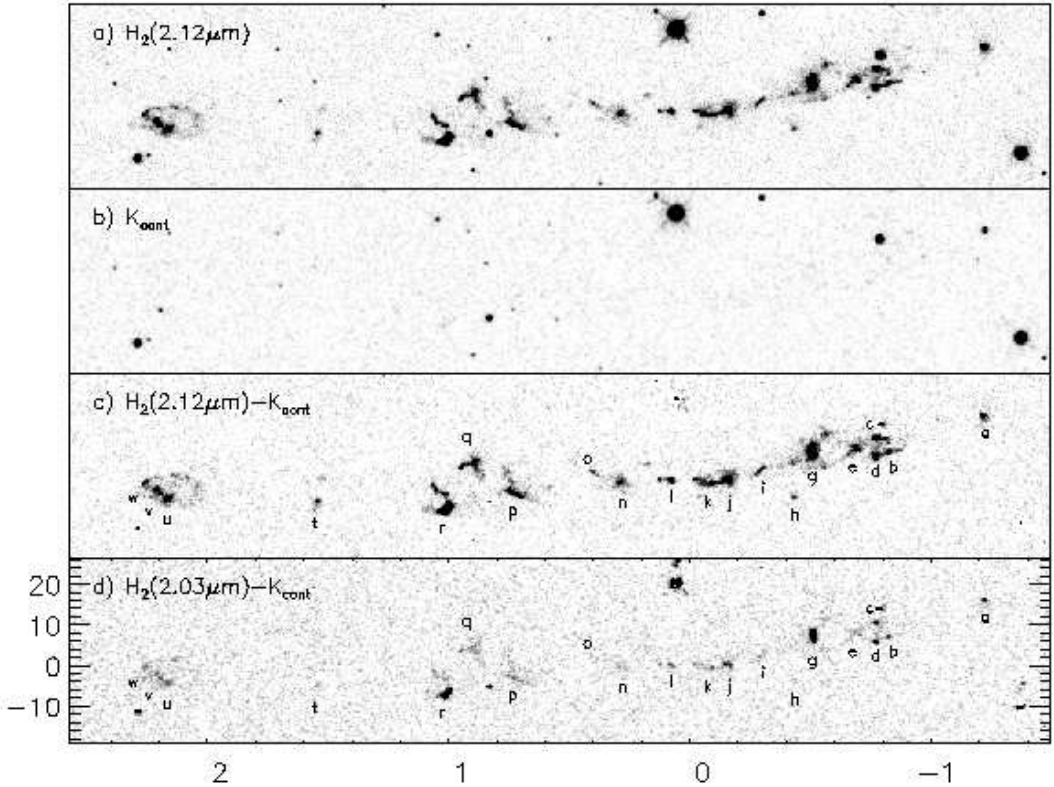}
\caption{ Near-infrared images of the outflow: 
a) H$_2$ 1--0 S(1) 2.122 $\mu$m line, 
b) K$_{\rm{cont}}$,
c) continuum subtracted H$_2$ 1--0 S(1) 2.122 $\mu$m,
d) continuum subtracted H$_2$ 1--0 S(2) 2.034 $\mu$m.
The horizontal scale is in arcminutes and the vertical
scale is in arcseconds. The origin is set at 
($\alpha$,$\delta$)(J2000) =
($06^{\rm{h}}35^{\rm{m}}25\fs 0$, $+03\arcdeg56\arcmin21\arcsec$)
}
\end{figure*}

\begin{deluxetable}{lccrrc}
\tablecaption{Positions and flux estimates for the NIR H$_{2}$ knots of 
MHO 1321}
\tablewidth{0pt}
\tablehead{\colhead{Knot} & \colhead{R.A. (J2000)} & \colhead{Dec (J2000)} &
\colhead{H$_2$ 1--0 S(2)} & \colhead{H$_2$ 1--0 S(1)} 
}\startdata
a  & 06:35:20.08  &  +03:56:37.9 &    4.1$\pm$2.0 &    7.6$\pm$2.2  \\
b  & 06:35:21.64  &  +03:56:28.8 &    2.2$\pm$1.4 &    8.9$\pm$2.2  \\
c  & 06:35:21.82  &  +03:56:32.3 &    3.9$\pm$1.9 &   12.5$\pm$2.4  \\
d  & 06:35:21.85  &  +03:56:27.6 &    4.2$\pm$1.8 &   15.6$\pm$2.8  \\
e  & 06:35:22.19  &  +03:56:29.5 &    2.2$\pm$1.7 &   10.1$\pm$2.4  \\
f  & 06:35:22.70  &  +03:56:33.2 &    2.0$\pm$1.6 &    4.9$\pm$2.1  \\
g  & 06:35:22.90  &  +03:56:28.7 &   15.5$\pm$3.8 &   52.4$\pm$5.8  \\
h  & 06:35:23.21  &  +03:56:17.3 &    1.2$\pm$1.5 &    3.3$\pm$1.8  \\
i  & 06:35:23.77  &  +03:56:24.1 &    1.8$\pm$1.6 &    5.2$\pm$1.9  \\
j  & 06:35:24.33  &  +03:56:21.6 &    4.7$\pm$2.2 &   20.8$\pm$3.4  \\
k  & 06:35:24.61  &  +03:56:20.7 &    2.5$\pm$1.6 &   10.4$\pm$2.4  \\
l  & 06:35:25.26  &  +03:56:21.4 &    2.4$\pm$1.6 &    8.3$\pm$2.1  \\
m  & 06:35:25.46  &  +03:56:21.6 &    1.5$\pm$1.3 &    4.3$\pm$1.8  \\
n  & 06:35:26.13  &  +03:56:20.9 &    2.0$\pm$1.7 &    8.6$\pm$2.5  \\
o  & 06:35:26.57  &  +03:56:23.5 &    1.4$\pm$1.4 &    3.3$\pm$1.6  \\
p  & 06:35:27.90  &  +03:56:18.3 &    3.1$\pm$1.9 &   11.8$\pm$2.6  \\
q  & 06:35:28.56  &  +03:56:25.7 &    3.0$\pm$2.0 &   10.8$\pm$2.6  \\
r  & 06:35:29.02  &  +03:56:14.2 &   12.8$\pm$3.4 &   42.3$\pm$5.0  \\
s  & 06:35:29.18  &  +03:56:18.6 &    2.1$\pm$1.6 &    6.9$\pm$2.0  \\
t  & 06:35:31.16  &  +03:56:15.8 &    1.6$\pm$1.5 &    4.3$\pm$1.8  \\
u  & 06:35:33.64  &  +03:56:16.6 &    3.0$\pm$1.7 &   11.3$\pm$2.6  \\
v  & 06:35:33.85  &  +03:56:18.9 &    2.8$\pm$1.9 &    7.9$\pm$2.1  \\
w  & 06:35:34.04  &  +03:56:20.5 &    2.0$\pm$1.5 &    5.1$\pm$2.0  
\enddata
\tablecomments{Flux is in units of $10^{-18}$ W m$^{-2}$. Aperture radius
used is 7 pixels. Flux uncertainty includes calibration uncertainty}
\end{deluxetable}

\subsection{IRAC color space of shocked gas}

In order to study the structure of the  shocked gas, we applied the IRAC
color analysis method developed by \citet{yba2009}. We improved the color
analysis method by including the effects of CO $\nu$=1--0 band emission in
the total emission of the shocked gas. The distribution of the
population of pure rotational levels of CO due to collisional
excitation with H$_2$, H, and He was calculated using the rate
coefficients of \citet{dra1984}. We employed the method of
\citet{gon2002} to calculate the relative rotational population for
the CO $\nu$=1 vibrational level. The Einstein A-values for the CO
$\nu$=1--0 rovibrational transitions were obtained using the
oscillator strengths of \citet{hur1993}. In our calculations, we set
$n_{\rm{H}} = n({\rm{H}}) +  2n({\rm{H}}_{2})$, 
$n({\rm{He}})/ n_{\rm{H}} = 0.1 $ and $n({\rm{CO}})/ n_{\rm{H}}=7
\times 10^{-5}$. The fraction of atomic to molecular hydrogen was
estimated by considering the rate of collisional dissociation
by H atoms, 
$R_{d} = 1.0 \times 10^{-10} \exp(-52\,000/T) \: {\rm{cm^{3}\,s^{-1}}}
$
\citep{leb2002} and the rate of formation on grains, 
$
R_{f} = 3.8 \times 10^{-17} T^{-1.5} \: {\rm{cm^{3}\,s^{-1}}}
$, derived from \citet{hol1979} with the cooling
rates for H$_2$ and H$_2$O \citep{leb1999,leb2002}.

Figure 3 shows the location of shocked gas in IRAC [3.6]--[4.5] versus
[4.5]--[5.8] color space for maximum shock temperature of
$T_{\rm{max}} = 6 \times 10^{3}$ K for gas temperatures $T =
1500-5000$ K and densities $n_{\rm{H}} = 10^{5}-10^{7}$
cm$^{-3}$. The square brackets refer to IRAC magnitudes.
The post-shock fraction of atomic hydrogen was found to be
$n({\rm{H}})/n_{\rm{H}} \sim 1-4 \times 10^{-3}$ which is consistent
with simulations of non-dissociative C-shocks \citep{wil2000}.
In the simulations by \citet{wil2000} it was found that the
atomic fraction is relatively constant over  a wide range of maximum
temperatures. Therefore we will assume our color space to be
representative of non-dissociative C-shocks in general.  
The location of the shocked emission in
IRAC color space depends on the gas density, fraction of atomic
hydrogen, and the kinetic temperature of the gas.  The [4.5]--[5.8]
color is strongly dependent on temperature, while the [3.6]--[4.5]
color has a strong dependence on the atomic hydrogen density. At high
densities, the dependence on density decreases as the H$_2$ gas moves toward
local thermal equilibrium (LTE). 
The slope of the reddening vector is similar to the approximate slope of 
the constant temperature lines. Thus temperature maps of high extinction
regions remain accurate even if the extinction cannot be accounted for. 
However, unless extinction can be corrected for, accurate density
information may not be attainable.

We fit an analytic form to the relationship between color and temperature 
for the non-dissociative case,
\begin{eqnarray}
T_{3} = 4.19 - 0.97([3.6]\!\!-\!\![4.5]) - 2.11([4.5]\!\!-\!\![5.8]) \nonumber\\
       +0.59([3.6]\!\!-\!\![4.5])^{2} + 0.50([4.5]\!\!-\!\![5.8])^{2} \nonumber
\end{eqnarray}
where $T_{3} = T/10^{3}$ in the color space defined by
$2.0 > [3.6]\!\!-\!\![4.5] > -0.21([4.5]\!\!-\!\![5.8]) + 1.5$, 
and $-0.2 < [4.5]\!\!-\!\![5.8] < 2.0$. 
The difference between the analytic fit and the calculated 
temperature-color relation over the defined range is less than 10\%. 
Additionally, one can use the flux in the 3.6 $\mu$m image to estimate
the column density of the shocked H$_2$.
Using our calculations we fit the following analytic form to the 
relationship between column density, IRAC 3.6 $\mu$m flux density, and temperature,
\begin{eqnarray}
\log(N_{\rm{H}_{2}}/F_{3.6}) = 23.11 - 3.40T_{3} + 0.742T_{3}^{2} \nonumber\\ 
 - 0.0589T_{3}^{3} - 0.071n_{6} + 0.012n_{6}T_{3} \nonumber 
\end{eqnarray} 
where $N_{\rm{H}_{2}}$ is the column density of shocked H$_2$ in cm$^{-2}$, $F_{3.6}$
is the IRAC 3.6 $\mu$m band flux density in units of MJy sr$^{-1}$, 
and $n_{6} = n_{\rm{H}}/10^{6}$
for $1.5 < T_{3} < 5.0$ and $2 < n_{6} < 10$.

\begin{figure}
\plotone{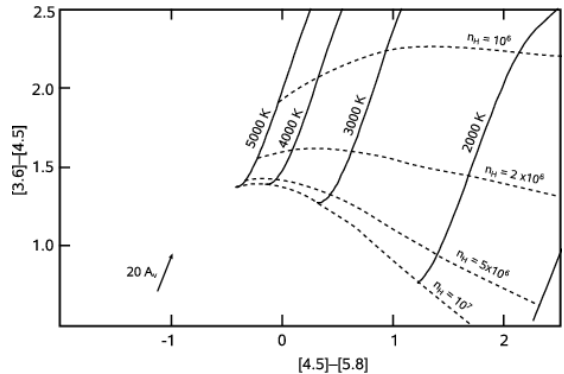}
\caption{ IRAC color-color plot indicating the region occupied by shocked 
gas composed of H$_2$ and CO for T$_{\rm{max}} = 6 \times 10^{3}$ K 
} 
\end{figure}

In order to investigate the color space of dissociative J-type
shocks we simulated the gas properties of a dissociative shock by 
setting the maximum
temperature of the gas to $T_{\rm{max}} = 1 \times 10^{4}$ K. 
Our calculations show significant dissociation of H$_2$ with increasing density. 
Figure 4 shows the IRAC color space
for dissociatively shocked gas. As the molecular hydrogen gets 
dissociated, emission from the CO $\nu$=1--0 band begins to 
dominate the 4.5 $\mu$m IRAC channel. 

The relationships between temperature, density and color are different
for the case of the non-dissociative shock and the case of the 
dissociative shock. 
There is some degeneracy at the low density and low dissociation region 
of color space for the dissociative shock and at the high density non-dissociative
shock region. These two models meet in a region of color space
defined by H$_2$ gas in LTE.
Although there is degeneracy, we expect the distribution in color
space for the dissociative shock to primarily lie at [4.5]--[5.8] $<$ 0.
Thus we define the domain of the dissociative shocked gas in 
color space to be [4.5]--[5.8] $< 0$ and [3.6]--[4.5] $> 1.5$,
whereas we define the color domain of non-dissociate gas to lie primarily 
at [4.5]--[5.8]$ > 0$. By analyzing the color distribution
of an outflow, it may be possible to distinguish between the 
cases. A pixel density distribution, produced from binning the 
colors of each pixel in the the outflow, can reveal the 
nature of the outflow by showing where most of the pixels 
lie in color space. 

It should be noted that the IRAC color analysis assumes dust and PAH emission is negligible. In order to prevent dust emission from affecting the color analysis, the 8 $\mu$m IRAC color is not used as there is evidence in some shocks of continuum dust emission within that wavelength range covered by the 8 $\mu$m channel \citep[eg.][]{smi2006}. Additionally, PAHs are very likely destroyed in shocks. In a recent study, \citet{mic2010} show that strong shocks can destroy PAHs or severely denature them. 

\begin{figure}
\plotone{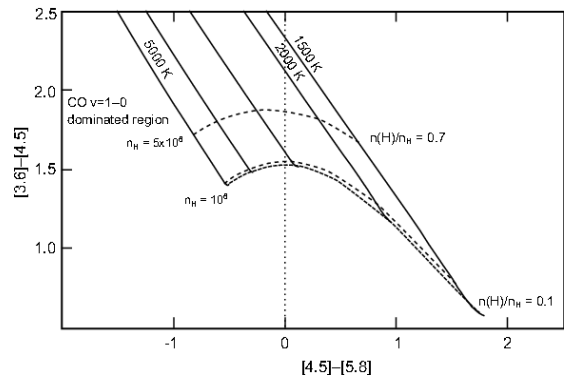}
\caption{ IRAC color-color plot indicating the region occupied by 
dissociatively shocked gas composed of H$_2$ and CO 
} 
\end{figure}

\subsection{IRAC color analysis of MHO 1321}

The IRAC 8 $\mu$m image of the outflow region reveals patches of absorption
against the diffuse background (Figure 1). Of particular interest is a dark patch seen 
in absorption that bisects the outflow. We created an extinction map from 
the 8$\mu$m data
using a small scale median filter assuming a uniform background.
We applied this extinction map to the images of
the outflow region using the mid-infrared reddening law (KP, v5.0)
of \citet{cha2009}. However, this is not able to 
account for the total extinction in the line of sight toward the outflow. 
Nonetheless, the temperature-color relation is insensitive to extinction 
for non-dissociative shocks.
We estimate the background using a ring median filter 
and subsequently remove this background from the IRAC 
3.6 $\mu$m, 4.5 $\mu$m, and 5.8 $\mu$m
images.
A ring median filter is a median filter from which only the pixels
within an annulus are used in calculating the median \citep{sec1995}. 
The scale of this filter needs to be larger than the scale of the 
shocked emission otherwise the background will be overestimated, yet
small enough to account for the large scale background fluctuations.
The images were shifted and
registered with each other and then IRAC colors at each pixel location
were determined. Figure 5 shows the pixel density in IRAC color space
for the knots of the outflow. Due to the lack of pixels whose colors
are in or near the CO dominated region and our criteria above for
non-dissociative shocks, we conclude that this shock is mostly
non-dissociative and we can therefore estimate the thermal structure
based on color analysis. We compared the colors to those of non-dissociative 
shocked gas with
the cutoff [4.5]--[5.8] $\leq$ 1.5. A thermal map was 
created by estimating the gas temperature based on the
location of the pixels in color space. Figure 6 shows the 
thermal map  of the outflow.

\begin{figure}
\plotone{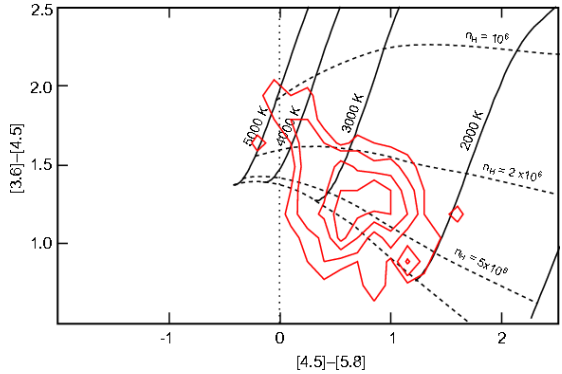}
\caption{Contours indicate the pixel density in IRAC color space of the 
outflow knots. The distribution of pixels and the lack of pixels in the CO 
dominated region indicate non-dissociative shocks.} 
\end{figure}

We find that most of the NIR H$_2$ knots are spatially coincident with
the high temperature regions of the flow.  However, knots a and v  do not
have a corresponding IRAC derived temperature. The NIR images reveal stars in the line of sight
for these knots which add to the emission and prevents IRAC color analysis from deriving
temperatures. 
Seven of the knots have estimated temperatures greater or equal to  $3 \times 10^{3}$ K. 
By combining the NIR infrared and temperature data it is possible to
estimate the extinction towards the brightest knots. For this we used the extinction 
cross-sections of \citet{wei2001}.  The median extinction to the knots is A$_{\rm{v}}$ = 27.  The
knots j and k in the vicinity of the dark clump have higher extinction
compared to the rest of the knots. We use the median extinction value
to de-redden the 3.6 $\mu$m flux and use it create a column density
map with our column density temperature relation. 
Figure 7 show the
column density map of shocked H$_2$ in the flow. We find that there is
also a correspondence between the NIR H$_2$ knots and regions of
higher column density. Using the established distance to the RMC of 1.6 kpc and the 
column density map we calculate the total mass of the shocked H$_2$ 
(T $> 2000$ K) in the 
outflow to be $3.5 \times 10^{30}$ g ($\sim 2 \times 10^{-3} M_{\sun}$).

\notetoeditor{Please print this figure across both columns and in color}
\begin{figure*}
\plotone{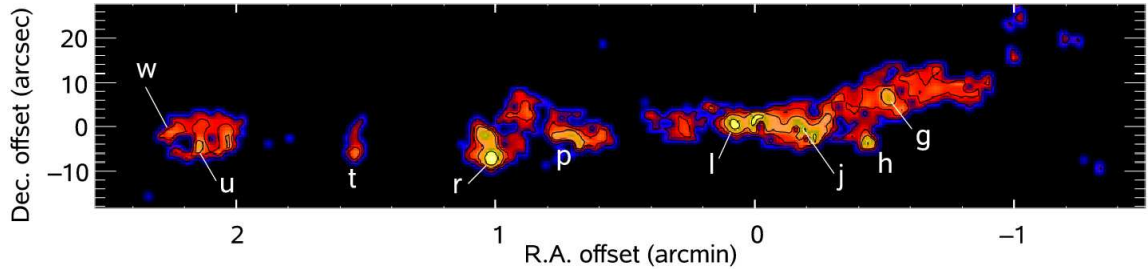}
\caption{Thermal map of the outflow based on color analysis of the IRAC 
data. The contour levels indicate T = 1500 K, 2500 K, 3000 K, 4000 K.
The origin is set at 
($\alpha$,$\delta$)(J2000) =
($06^{\rm{h}}35^{\rm{m}}25\fs 0$, $+03\arcdeg56\arcmin21\arcsec$)
}
\end{figure*}

\notetoeditor{Please print this figure across both columns and in color}
\begin{figure*}
\plotone{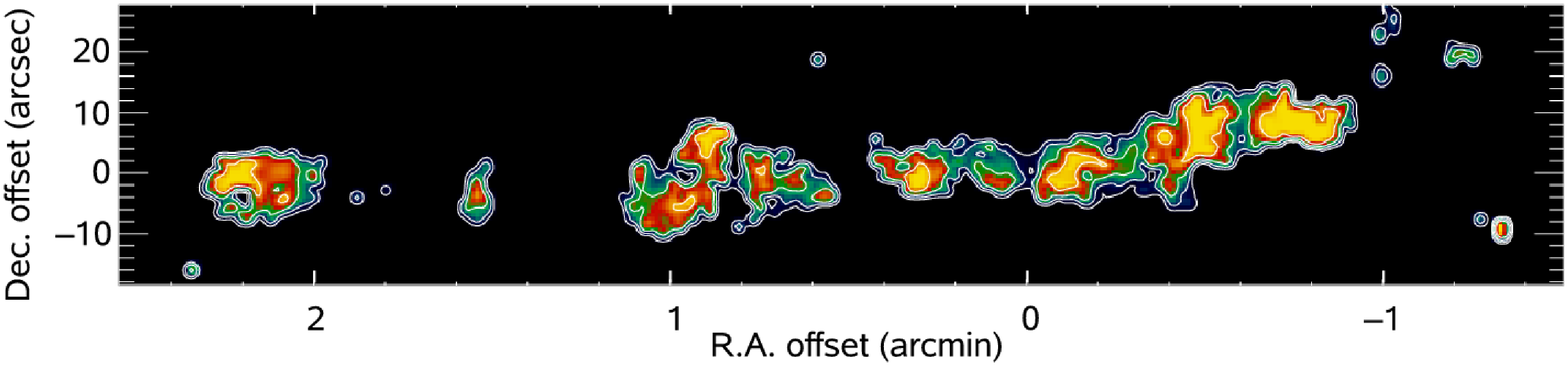}
\caption{Column density map for H$_2$ of the outflow based on color 
analysis of the IRAC 
data. The contour levels indicate 
$N_{\rm{H}_{2}}$ = 
$2\times10^{17}$ cm$^{-2}$, 
$5\times10^{17}$ cm$^{-2}$,
$1\times10^{18}$ cm$^{-2}$,
$2\times10^{18}$ cm$^{-2}$.
The origin is set at 
($\alpha$,$\delta$)(J2000) =
($06^{\rm{h}}35^{\rm{m}}25\fs 0$, $+03\arcdeg56\arcmin21\arcsec$)
}
\end{figure*}

\begin{deluxetable}{lrrr}
\tablecaption{IRAC color analysis of H$_2$ knots}
\tablewidth{0pt}
\tablehead{\colhead{Knot} & \colhead{T (10$^{3}$ K)} & 
\colhead{N$_{\rm{H2}}$ (10$^{18}$ cm$^{-2}$)}
} 
\startdata
b &  2.5$\pm$0.2 &  3.7$\pm$0.3\\ 
c &  2.7$\pm$0.3 &  3.7$\pm$0.2\\
d &  2.5$\pm$0.2 &  4.1$\pm$0.2\\
e &  2.5$\pm$0.2 &  2.2$\pm$0.2\\
f &  2.5$\pm$0.2 &  2.6$\pm$0.2\\
g &  3.2$\pm$0.4 &  3.0$\pm$0.1\\
h &  3.0$\pm$0.7 &  0.6$\pm$0.1\\
i &  3.0$\pm$0.2 &  1.1$\pm$0.1\\
j &  3.3$\pm$0.6 &  2.1$\pm$0.1\\
k &  2.8$\pm$0.3 &  2.3$\pm$0.2\\
l &  3.3$\pm$0.8 &  1.3$\pm$0.1\\
m &  2.5$\pm$0.3 &  1.1$\pm$0.2\\
n &  2.3$\pm$0.2 &  2.9$\pm$0.4\\
o &  2.3$\pm$0.1 &  1.5$\pm$0.3\\
p &  2.8$\pm$0.6 &  2.0$\pm$0.2\\
q &  2.6$\pm$0.1 &  2.1$\pm$0.2\\
r &  4.0$\pm$0.8 &  1.4$\pm$0.1\\
s &  2.8$\pm$0.6 &  1.0$\pm$0.2\\
t &  2.6$\pm$0.3 &  1.0$\pm$0.2\\
u &  3.3$\pm$0.2 &  1.4$\pm$0.1\\
w &  2.7$\pm$0.2 &  2.4$\pm$0.2\\
\enddata
\tablecomments{The estimated temperature is the column density averaged
  temperature determined from pixel colors within each aperture. The
  temperature uncertainty is the column density weighted standard deviation
  of the temperatures corresponding to the individual pixels. }
\end{deluxetable}

\subsection{Outflow Source}

The source of the outflow is not seen in the NIR nor in the IRAC images.
However, inspection of the MIPS 24 $\mu$m image reveals a source
(($\alpha$,$\delta$)(J2000) =
($06^{\rm{h}}35^{\rm{m}}25\fs 0$, $+03\arcdeg56\arcmin21\arcsec$))
bisecting
the outflow (Figure 8). Moreover, this source is spatially coincident with a 
dark patch seen in the 8 $\micron$ image. 
This patch is elongated nearly perpendicular to the outflow and the northwest part of it 
has the morphology of an outflow cavity. 
The dark patch is seen in the contours that indicate mass surface densities  
obtained through extinction mapping of the 8 $\mu$m imaging data by the method 
of \citet{but2009}. The contour levels correspond to mass surface densities of
$\Sigma$ = (2.5, 4.0, 5.0, 6.0) $\times$ 10$^{-3}$ g cm$^{-2}$. 
This small cloud may be a remnant of the core from which the protostar formed.  
The morphology of the 
northern end of the cloud appears as a bi-polar outflow cavity with an 
opening angle $\theta \sim 30^{\circ}-40^{\circ}$. 
H$_2$ 1--0 S(1) surface brightness contours are overlaid on the MIPS 24 $\mu$m 
image that bisect the MIPS source and a coincident with the cavity of 
the dark cloud. 
This source is also detected in the MIPS 70 $\mu$m 
and MIPS 160 $\mu$m imaging data.
However, we are unable to estimate the flux in the MIPS 70 $\mu$m band image 
due to  incomplete coverage and possible contamination 
from the adjacent knot (j) which may 
have emission from dust and the 63 $\mu$m [\ion{O}{1}] line \citep{rei2001}.
Similarly, the MIPS 160 $\mu$m band image may include emission
from the knot in addition to the source. 
This source is not detected at shorter wavelengths and thus
can be classified as a Class 0 protostar. 
We propose this newly discovered protostar to be the source of the outflow. 
Additionally, using the column density map, we find that the mass of the 
shocked H$_{2}$ to the east 
($1.8 \times 10^{30}$ g) of this source is almost equal 
to the mass west of the source ($1.7 \times 10^{30}$ g) .

Faint extended 24 $\mu$m emission is also detected at the locations of the 
brightest H$_2$ knots (g \& j). 
This may arise from fine-structure [\ion{Fe}{2}] lines 
within the MIPS 24 $\mu$m band \citep{vel2007}.  This is consistent with our IRAC color analysis 
of these knots that reveal them to be high temperature regions ($T \geq 3.2 \times 10^{3}$ K).
This consistency between the IRAC color analysis and the MIP 24 $\mu$m emission validates 
our usage of the color space for non-dissociative shocks.

\begin{figure}
\plotone{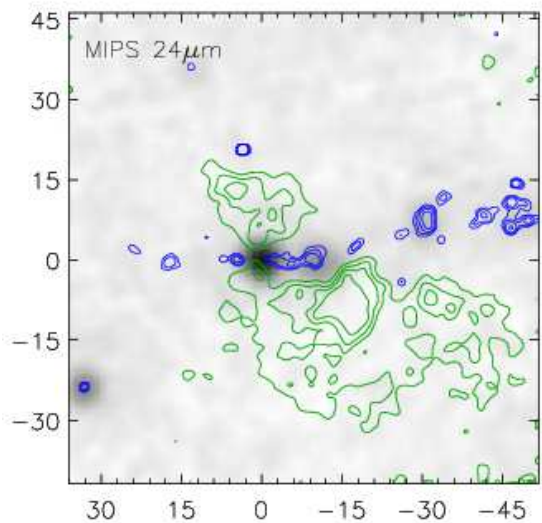}
\caption{ MIPS 24 $\mu$m image of the outflow sources. The green contours 
show the dark cloud that bisect the outflow and indicate
the $\Sigma$ values (2.5, 4.0, 5.0, and 6.0) $\times 10^{-3}$ g cm$^{-2}$
obtained through extinction mapping using the IRAC 8 $\mu$m imaging data \citep{but2009}. 
The blue contours are H$_2$ 1$-$0 S(1) surface brightness contours
with levels indicating (0.5, 1.0, 2.0, 5.0) $\times 10^{-18}$ W m$^{-2}$ 
sr$^{-1}$. The H$_2$ contours reveal the location of the outflow.
The scale of the image axes is in arcseconds. 
The origin is set at 
($\alpha$,$\delta$)(J2000) =
($06^{\rm{h}}35^{\rm{m}}25\fs 0$, $+03\arcdeg56\arcmin21\arcsec$)    
}
\end{figure}

\section{Discussion}

\subsection{Structure of the outflow}

The long axis of the flow extends to 3.3$'$. With a distance of 1.6
kpc to the RMC, the flow would have a projected total length of 1.5 pc. The east
lobe extends 2.3$'$ from the MIPS source to the bow shock, 
while the west lobe extends only 1$'$. Assuming a projected outflow velocity of
100 km s$^{-1}$, using the east lobe we estimate the age of the outflow to be  
$~ 10^{4}$ years.  
This age is consistent with the typical age of a Class 0 source and thus 
this outflow may provide an accretion record of the protostar \citep{rei2001}.
We can estimate the mass flux of the outflow as
$\dot{M} \sim M({\rm{H}}_{2}) v_{t} l_{t}$ where $v_{t}$ is the projected outflow velcocity
and  $l_{t}$ is the projected outflow length. Assuming the typical value of $v_{t} = 100$ km s$^{-1}$ and using the values for the H$_2$ mass and length of the east lobe, we
estimate a mass flux of $\dot{M} \sim 10^{-7} \, {\rm{M}}_{\sun} \,  {\rm{yr}}^{-1}$ . 
This value is consistent with those
obtained from other outflows using spectroscopic data \citep{pod2006}.  

The NIR H$_{2}$ data reveals the higher temperature
regions of the outflow seen in the IRAC images. The IRAC images also
show the cooler regions of the flow as the IRAC bands contain pure
rotational H$_2$ lines in addition to $\nu$ = 1--0 and $\nu$ = 2--1
ro-vibrational lines.

The flow is spatially coincident with two lobes of high velocity CO gas
observed by \citet{den2009}.  Similar to many HH flows which extend 
further than their CO
counterpart, we find the eastern half to extend beyond the the
eastern CO lobe. 
The east end of the H$_2$ flow ends in a large bow
shock, while the west end reveals a complex structure resembling
either a broken up bow shock or multiple smaller bow shocks. 
Although the outflow is linear on large scales, the IRAC and
H$_2$ data reveal a region in the eastern lobe before the bow shock
with a more chaotic structure.
This deviation from a linear progression of knots may be due to
a possible interaction with another outflow. The CO observations 
of \citet{den2009} reveal another flow in the NE-SW direction originating
from the embedded cluster PL07 \citep{phe1997} that points toward this region. 
A collision between the two flows may explain the morphology and 
high temperature of knot r.
The distribution of knots may also be due to variations in in jet direction 
over time and thus an indication of jet precession. 

\subsection{Deflection of the Outflow}

The western end of the outflow appears slightly bent northward. 
The outflow is deflected by an angle $\theta_{d} = 20^{\circ}$
where it appears to graze the densest region within the dark
patch. The
deflection angle remains small and appears to decrease slightly beyond
the interaction region. This is consistent with the simulations by
\citet{bae2009} of outflows colliding with dense cloud cores where
the impact parameter is large. The deflection of the outflow may
explain why the western end of the flow is shorter than the eastern
end as the outflow velocity is expected to decrease after the collision
\citep{rag1995}. This is consistent with the IRAC color analysis that
reveals high temperature shocked gas (knot j)  to the east of the collision.

If the outflow is composed of episodic ejection of material, there may
be collision between clumps of material moving through the flow due 
to the velocity change \citep{rag2003}. As these successive clumps collide they 
may give rise to the high temperature and high column density region (g) 
found slightly west of the deflection.
This interaction may also explain why the western lobe lacks the bow shock 
structure seen in the eastern end.    



\section{Conclusions}

We present the discovery of a new bi-polar outflow in the 
Rosette Molecular Cloud and use NIR narrowband
and Spitzer imaging data to study the flow. We show that 
IRAC color analysis can be used to interpret the interaction of an outflow with its surrounding environment.
Using our calculations of the IRAC space of non-dissociative shocked gas we fit analytic forms to the color-temperature and column density-temperature relationships. We verify that IRAC color analysis
can reveal regions of shocked gas and find that the NIR H$_{2}$ knots
correspond to regions of high temperature and or column density determined through color
 analysis.  
 We find diffuse MIPS 24 $\mu$m emission, most likely from [\ion{Fe}{2}] lines, to be coincident 
 with regions of high temperature thus
 confirming the validity of using the 
 non-dissociative shock IRAC color space
 The NIR line ratios combined with the temperature estimates allow for the
 determination of extinction along the line of sight which is used to create
 a column density map of the shocked H$_{2}$ gas. 
 We deduce that the asymmetry in the outflow is due to interactions with the dense
material to the west of the outflow source causing deflection and possibly deceleration
of the outflowing material.

\acknowledgments

We thank Mike Butler for running his extinction mapping code on the IRAC data
and providing the mass surface density data shown in Figure 8. 
We thank the CTIO Blanco-4m observatory support scientists and staff 
including Nicole van der Bliek,
Hernan Tirado, and Alberto Alvarez.
We thank our referee, John Bally, for his useful comments and
suggestions. 
This work is based in part on archival data obtained with the Spitzer
Space Telescope, which is operated by the Jet Propulsion Laboratory,
California Institute of Technology under a contract with NASA. Support
for this work was provided by an award issued by JPL/Caltech and also
a NASA LTSA Grant NNG05GD66G. JY acknowledges support by a
Florida Space Grant Fellowship from NASA through the 
Florida Space Grant Consortium.



{\it Facilities:} \facility{Spitzer (IRAC,MIPS)}, \facility{Blanco (ISPI)} .



\clearpage

\end{document}